\documentclass[a4paper,prl,twocolumn,superscriptaddress,longbibliography]{revtex4-1}
\usepackage{amsmath}
\usepackage{amssymb}
\usepackage{graphicx,rotating,subfigure,amsmath,amsfonts,amssymb,delarray,color}
\usepackage{amsthm}
\usepackage{color}
\usepackage{comment}
\usepackage{float}
\usepackage{soul}

\let\<\langle
\let\>\rangle

\def\({\left(}
\def\){\right)}
\def\ll{\left\{}
\def\rr{\right\}}

\newcommand{\ket}[1]{|{#1}\rangle}
\newcommand{\bra}[1]{\langle{#1}|}

\newcommand{\up}{\uparrow}
\newcommand{\down}{\downarrow}

\begin{document}

\title{Signatures of many-body localization and metastability by weak perturbation}
\author{Yang Zhao}
\address{Department of Applied Physics, School of Natural and Applied Sciences, Northwestern Polytechnical University, Xi'an 710072, China}
\address{Asia Pacific Center for Theoretical Physics, Pohang 37673, Korea}
\author{Rajesh Narayanan}
\address{Department of Physics, Indian Institute of Technology Madras, Chennai 600036, India}
\author{Jaeyoon Cho}
\address{Asia Pacific Center for Theoretical Physics, Pohang 37673, Korea}
\address{Department of Physics, POSTECH, Pohang 37673, Korea}
\date{\today}
\begin{abstract}
	Nonequilibrium dynamics in isolated quantum many-body systems displays a number of intriguing features, such as many-body localization (MBL) and prethermalization.
	Here we investigate a simple ladder system with disorder, in which various distinct dynamical features coexist and interplay.
	By exact diagonalization, we demonstrate that the system exhibits the signatures of an MBL-ergodic-MBL reentrant transition, metastability, and disorder-free MBL.
	We give an account of these properties by introducing a quasi-particle picture and interpreting the quasi-vacuum energy fluctuation as an effective disorder on the quasi-particle dynamics.
	It is speculated that the weak perturbation behavior is a finite-size effect, but its relaxation time scale increases with the system size.
\end{abstract}
\maketitle


\textit{Introduction.---}%
Integrals of motion play a crucial role in nonequilibrium quantum dynamics. 
In general, even if the unitary evolution retains the vanishing entropy of the whole system in a pure state, subsystems tend to maximize the entropy under the constraint set by the integrals of motion, thereby approaching the generalized Gibbs ensemble~\cite{jay57,rig07,mor18}. 
In nonintegrable models lacking conserved quantities except global ones like energy and particle number, the stationary state then naturally reduces to the conventional Gibbs ensemble. 

This reasonable scenario is, however, defied by (nonintegrable) disordered systems exhibiting many-body localization (MBL), which do not thermalize into the Gibbs ensemble~\cite{bas06,oga07,zni08,pal10,hus13,kja14,nan15,bar12,vos13,and14,imb16,imb16b,sch15,bor16,cho16,hua17}. 
Although a thorough understanding of MBL is still lacking, growing evidence has led to a general agreement on the physics of MBL in one-dimensional (1d) disordered Heisenberg chains and similar systems.
In such systems, while weak disorder thermalizes the system, strong disorder does not.
The transition between the two different behaviors is believed to be a phase transition~\cite{pal10,hus13,kja14,nan15}. 
In the case of potential disorder, the MBL is signified by a Poissonian level statistics and a $\log(t)$ scaling of entanglement entropies in time $t$~\cite{zni08,bar12}.
In the case of bond disorder, the dynamical renormalization group analysis predicts that the entanglement entropy scales as $\log^{2/\phi}(t)$ with $\phi$ the golden ratio~\cite{vos13,hua17}.

While MBL in higher dimension is still open to question~\cite{der17}, it is remarkable that the MBL physics already becomes richer only by a slight generalization, namely, by considering a ladder system.
For example, one can consider a situation where fast-moving particles are localized even in the absence of disorder by an effective potential produced by slow-moving heavy particles, although the time scale and the finite-size effect should be treated carefully~\cite{pap15,yao16}.
In fact, the finite-size effect is an ever present bugbear even in conventional (disordered) and well-studied MBL systems (see, e.g., Ref.~\cite{sou17}).
There are also disorder-free localization models in which a spin environment produces an effective disorder on a fermion dynamics~\cite{smi17a,smi17b}.

Another intriguing phenomenon arises when the system is nonintegrable but close to integrable.
Such a weak breaking of integrability typically sets in a separation of time scales, so that a metastable, prethermal period emerges before the ultimate thermalization~\cite{mar13,ess14,ber15,mor18}.
Note that in general, this phenomenon is irrelevant to MBL: whereas prethermalization emerges when integrability is broken by weak enough perturbation (see Ref.~\cite{lan18} for different possibilities), MBL emerges when it is broken by strong enough disorder.

In this paper, we demonstrate a simple ladder system in which above-mentioned, seemingly distinct, dynamical features coexist and interplay.
The system we consider is a ladder composed of two transverse-field Ising models coupled along the rungs by a coupling in the transverse direction (see Eq.~\eqref{eq:orig}).
This system has the following properties. 
First, it undergoes a reentrant transition in that the quintessential signatures of MBL---the Poissonian level statistics and the $\log(t)$ scaling of the entanglement entropy---are shown when the disorder is weak or strong, while it thermalizes when the disorder is of intermediate strength.
Second, when the disorder is weak, several distinct time scales emerge and a long metastable period appears, as in the case of prethermalization.
However, an important difference is that in our system, the dynamics follows the $\log(t)$ scaling behavior apart from the plateaus and saturates at a nonthermal state.
Third, the $\log(t)$ behavior of MBL persists even in the absence of disorder
when the disorder is replaced by a constant weak perturbation.
We give an account of these properties by introducing a quasiparticle picture.
Regarding the thermodynamic limit, our results, based on exact diagonalization, {\em suggest} that the slow dynamics by weak perturbation appears to be a finite-size effect.
However, there is a trade-off in that as the system size increases, while the slow dynamics is more pronounced with a longer relaxation time scale, the parameter range for its observation becomes narrower.
We could not arrive at a firm conclusion on this point.


\textit{Model.---}%
We consider the following Hamiltonian:
\begin{equation}
H = \sum_{i=1}^{L/2} \ll
\sum_{\alpha=1}^2 \(
h_i \sigma_{i,\alpha}^x
- J \sigma_{i,\alpha}^z \sigma_{i+1,\alpha}^z
\)
- g \sigma_{i,1}^x \sigma_{i,2}^x
\rr,
\label{eq:orig}	
\end{equation}
where $\sigma_{i,\alpha}^{x,z}$ denotes the Pauli operator acting on the $i$-th site of the $\alpha$-th ladder and the open boundary condition $\sigma_{L/2+1,\alpha}^{z}=0$ is considered.
We introduce disorder by taking $h_i$ randomly and uniformly from $[-D/2,D/2]$ with $D$ the disorder strength.
Note that $h_i$ is independent of $\alpha$.
This brings additional symmetry into the system, allowing us to simulate larger systems.

As the first step, we perform a duality transformation to the Hamiltonian~\cite{koh81} (see Appendix), which leads to
\begin{equation}
\begin{split}
	H = \sum_{i=1}^{L/2} & \ll
		h_i \(
			\sigma_{2i-1}^x \sigma_{2i}^x
			+ \sigma_{2i-1}^y \sigma_{2i}^y
		\)
		+ g \sigma_{2i-1}^z \sigma_{2i}^z
	\right. \\
	& \left.
		- J \(
			\sigma_{2i}^x \sigma_{2i+1}^x
			+ \sigma_{2i}^y \sigma_{2i+1}^y
		\)
	\rr.
\end{split}
\label{eq:hamil}
\end{equation}
This Hamiltonian reveals the symmetry of the system more explicitly.
The apparent one is the $U(1)$ symmetry described by the quantum number $\sum_i\sigma_i^z$.
Hereafter, we will consider the block of $\sum_i\sigma_i^z=0$.
Within this block, there exists another $Z_2$ symmetry described by $\prod_i\sigma_i^x$.
In order to obtain the energy level statistics, all these symmetries should be taken into account.

Hereafter, our discussion is based on Hamiltonian~\eqref{eq:hamil}.
In all the figures, each average curve is obtained by taking at least $10^3$ disorder samples, except for the cases of $L=16$ for which at least $10^2$ samples are taken, and the open boundary condition is taken unless stated otherwise.
We remark that as the periodic boundary condition is incompatible with the duality transformation, the periodic boundary cases of Hamiltonian~\eqref{eq:orig} and \eqref{eq:hamil} are different.

\begin{figure}
\begin{center}
\includegraphics[width=0.4\textwidth]{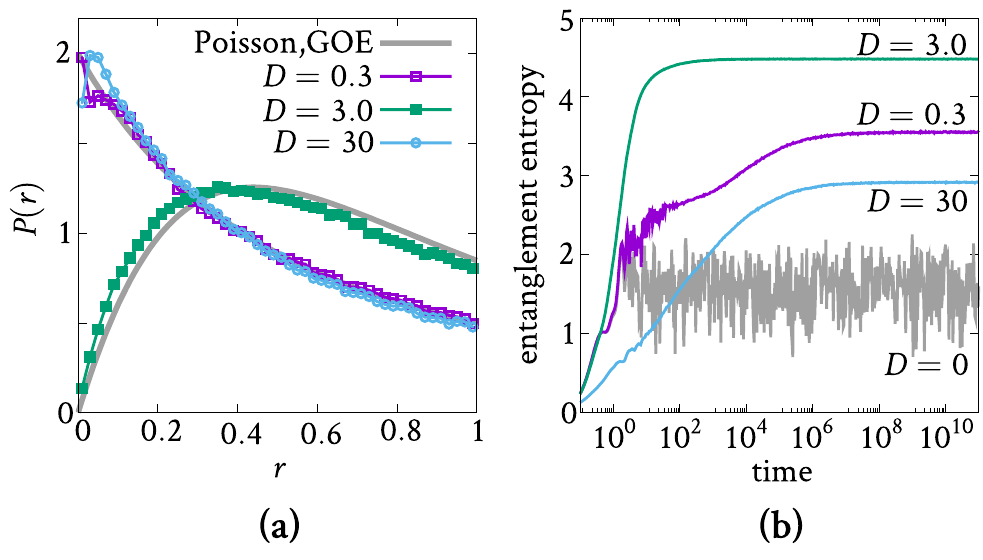}
\end{center}
\vspace{-1em}
\caption{(a) Energy level statistics and (b) average entanglement dynamics for $L=12$ and $J=g=1$.
For (b), the initial state is a Neel state for Hamilonian~\eqref{eq:hamil}.
}
\label{fig:reentrant}
\end{figure}

\begin{figure*}
	\begin{center}
		\includegraphics[width=0.96\textwidth]{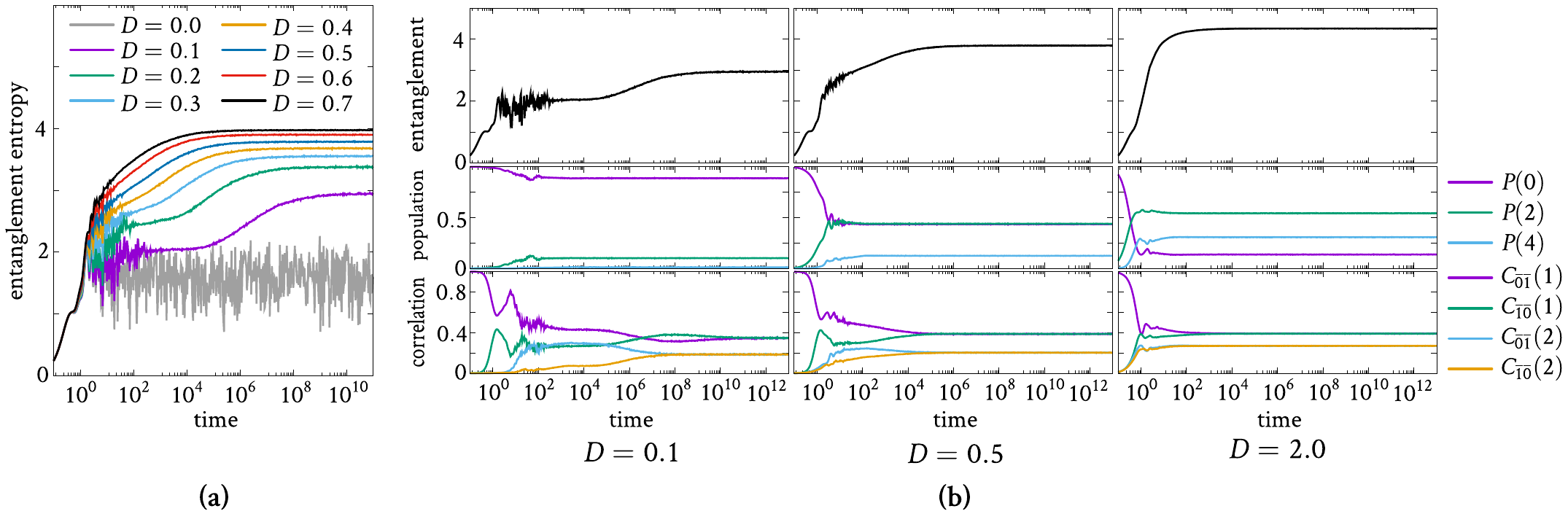}%
	\end{center}
	\vspace{-1em}	
	\caption{Average time evolution of the half-chain entanglement entropy, $P(2n)$, and $C_{\bar x\bar y}(d)$ for $L=12$ and $J=g=1$. 
	}
	\label{fig:scale}
\end{figure*}


\textit{Reentrant many-body localization transition.---}%
Our preliminary numerical results are shown in Fig.~\ref{fig:reentrant}.
Fig.~\ref{fig:reentrant}(a) shows the energy level statistics for three choices $D=\{0.3, 3.0, 30\}$ while keeping $J=g=1$ and $L=12$.
Here, we follow the convention in Ref.~\cite{oga07}, wherein the level statistics is obtained from the set of data 
$r_n={\min(\delta_n,\delta_{n-1})}/{\max(\delta_n,\delta_{n-1})}$
defined using energy eigenstates $E_n$ and level spacing $\delta_n=E_n-E_{n-1}$.
{The GOE curve is based on the values in Ref.~\cite{ata13}.}
In Fig.~\ref{fig:reentrant}(b), the average time evolution of the half-chain entanglement entropy is shown.
Here, the initial state is the Neel state $\ket{\down\up\cdots\down\up}_{1\cdots L}$, where $\ket\uparrow_i$ and $\ket\downarrow_i$ are the eigenstates of $\sigma_i^z=\ket\uparrow_i\bra\uparrow-\ket\downarrow_i\bra\downarrow$.
For $D=0.3$ and $D=30$, the level statistics is very close to the Poissonian and the half-chain entanglement entropy shows a $\log(t)$ scaling behavior.
These two features are the main signatures of MBL.
On the other hand, the ergodic signatures---GOE level statistics and a quick saturation of the entanglement entropy---are observed for the intermediate choice $D=3.0$.
Such a reentrant transition { with respect to the disorder strength} is unusual in the study of MBL.
It is natural to expect that the MBL is seen for large $D$.
In this case, one can regard each pair of spins at odd bond $(2i-1)-(2i)$ as a single spin of dimension 4, and interpret Hamiltonian~\eqref{eq:hamil} as if the first and the last terms (with $h_i$ and $J$) represent on-site disorder and nearest-neighbor interaction, respectively.
Then, the system reduces to a spin chain with strong on-site disorder.
Starting from this, it is expected to see the ergodic behavior when $D$ is decreased.
The nontrivial part here is the case of small $D$ in view of the fact that in conventional MBL, the localization appears only when the disorder is strong enough to overcome the effect of interaction.
In what follows, we mainly focus on this small $D$ case in order to uncover its nature.


\textit{Quasiparticle picture.---}%
Let us define for each even bond $(2i)-(2i+1)$,
\begin{equation}
\ket0_i = \ket{\up\down}_{2i,2i+1},~
\ket1_i = \ket{\down\up}_{2i,2i+1},
\end{equation}
where $\ket\up_{L+1}= \ket\up_1$ and $\ket\down_{L+1}= \ket\down_1$.
We also define the corresponding Pauli operators
\begin{equation}
X_i=\ket0_i\bra1+\ket1_i\bra0,~
Z_i=\ket0_i\bra0-\ket1_i\bra1.
\end{equation}
Similarly, we define
\begin{gather}
	\ket{\bar 0}_i = \ket{\up\up}_{2i,2i+1},~
	\ket{\bar 1}_i = \ket{\down\down}_{2i,2i+1},\\
	\bar X_i=\ket{\bar0}_i\bra{\bar1}
	+\ket{\bar 1}_i\bra{\bar0},~
	\bar Z_i=\ket{\bar0}_i\bra{\bar0}
	-\ket{\bar1}_i\bra{\bar1}.
\end{gather}
We will call $\ket{\bar0}$ and $\ket{\bar1}$ ``quasiparticles'', the motivation of which will be elaborated below.
The total number of quasiparticles can then be defined as $\bar N =\sum_i {\bar Z}_i^2$.
Note that in the subspace we consider ($\sum_i \sigma_i^z = 0$), the number of $\ket{\bar 0}$ and $\ket{\bar1}$ are always the same.
The entire Hilbert space $\mathcal{H}$ is thus split into $\mathcal H=\bigoplus_{n=0}^{L/4}\mathcal H_{2n}$ with $\mathcal H_{2n}$ being the subspace with $\bar N=2n$.
By abuse of terminology, we will call $\mathcal H_0$ the ``vacuum subspace''.

For $h_i=0$, Hamiltonian~\eqref{eq:hamil} becomes
\begin{equation}
\begin{split}
	H_0 = \sum_{i=1}^{L/2-1} & (
		- 2J X_i - g Z_i Z_{i+1} \\
		& - g Z_i \bar Z_{i+1}
		+ g \bar Z_i Z_{i+1}
		+ g \bar Z_i \bar Z_{i+1}
	).
\end{split}
\label{eq:quasi}
\end{equation}
Note that this Hamiltonian conserves $\bar N$.
Consequently, the dynamics in each Hilbert space $\mathcal{H}_{2n}$ is separated from the others.
For non-zero $h_i$, Hamiltonian~\eqref{eq:hamil} induces the following transitions:
\begin{equation}
\begin{cases}
	\ket0_i\ket0_{i+1}\leftrightarrow\ket{\bar0}_i\ket{\bar1}_{i+1},~
	\ket1_i\ket1_{i+1}\leftrightarrow\ket{\bar1}_i\ket{\bar0}_{i+1} \\
	\hfill (\mathcal H_{2n} \leftrightarrow \mathcal H_{2n+2}),\\
	\ket0_i\ket{\bar0}_{i+1}\leftrightarrow\ket{\bar0}_i\ket1_{i+1},~
	\ket1_i\ket{\bar1}_{i+1}\leftrightarrow\ket{\bar1}_i\ket0_{i+1} \\
	\hfill (\mathcal H_{2n} \leftrightarrow \mathcal H_{2n}).
\end{cases}
\label{eq:transition}
\end{equation}
The first transitions create or annihilate the two different types of quasiparticles in pairs.
The second transitions move the quasiparticles while altering the vacuum.
{Note that these transitions are associated with the creation, annihilation, and moving of ``holes'' in the vacuum, which change the vacuum energy.}


{
The above quasiparticle picture leads us to the following interpretation of the results for small $D$.
For example, consider subspace $\mathcal{H}_2$.
The state therein can be written as $\ket{\Psi(t)}=\sum_{i\not=j} c_{ij}(t)\ket{\bar0}_i\ket{\bar1}_j\ket{\phi(t)_{ij}}$, where $\ket{\phi(t)_{ij}}$ is the associated (normalized) vacuum state.
In a rotating frame (see Appendix), one can derive an effective quasiparticle Hamiltonian in the basis $\{\ket{\bar0}_i\ket{\bar1}_j\}$, where the diagonal elements are given by $\bra{\phi(t)_{ij}}H_0\ket{\phi(t)_{ij}}$ and the off-diagonal by transition~\eqref{eq:transition}.
Note that for $J=g=1$, the initial state $\otimes_{i=1}^{L/2}\ket0_i$ is a superposition of a broad range of energy eigenstates in the vacuum subspace and thus $\ket{\phi(t)_{ij}}$ evolves in a complicated manner depending on the quasiparticle position $\{i,j\}$.
As a result, the quasiparticle Hamiltonian has terms $\bar{Z}_i$, $\bar{Z}_i\bar{Z}_{i+1}$, and terms coming from transition~\eqref{eq:transition}, the coefficients of which are all effectively random.
A subtle issue here is that the effective disorder is time-dependent.
In general, dynamic disorder breaks localization~\cite{mad77}.
This rule does not apply in our case, however, as the effective disorder has an {\em infinite} correlation time.
A further discussion of this point is presented in Appendix.
}

Note that this effective disorder is independent of the original disorder in $h_i$.
This suggests that the present model may exhibit MBL behaviors even in the absence of disorder~\cite{pap15,yao16}.
This is indeed the case, as will be shown later.


\textit{Time scales.---}%
As the vacuum Hamiltonian (the first two terms in Eq.~\eqref{eq:quasi}) is integrable and broken by a small perturbation, a separation of time scales emerges~\cite{ber15,mor18}.
In Fig.~\ref{fig:scale}(a), we plot the average time evolution of the half-chain entanglement entropy for various values of $D$.
It can be seen that for small $D$, there are two plateaus in the entanglement dynamics.
Let us define three characteristic time scales in such a way that the first plateau begins at $t\simeq t_1$ and ends at $t\simeq t_2$, and then the second plateau begins at $t\simeq t_3$.
In order to examine these time scales, we define two quantities.
First, we define $P(2n)$ as the population in $\mathcal{H}_{2n}$.
Due to the aforementioned constraint, we have 
$\sum_{n=0}^{L/4} P(2n)=1$.
The second quantity is a correlation function
\begin{equation}
C_{\bar x\bar y}(d)
=
\bra{\Psi(t)}\left(
\frac{2}{\bar N}\sum_{i=1}^{L/2} \ket{\bar x}_i\bra{\bar x}\otimes
\ket{\bar y}_{i+d}\bra{\bar y}
\right)\ket{\Psi(t)},
\end{equation}
where $x,y\in\{0,1\}$ and the site index is defined modulo $L/2$.
By definition, $C_{\bar x\bar y}(d)=C_{\bar y\bar x}(L/2-d)$.
In Fig.~\ref{fig:scale}(b), we plot these two quantities along with the entanglement entropy.

Let us first discuss the dynamics for $D=0.1$ in Fig.~\ref{fig:scale}(b).
During the initial transient period from $t=0$ to $t_1$, quasiparticles are created from the vacuum.
In a short time, the entanglement entropy here shows a $\log(t)$ scaling behavior on average.
This feature can be seen more clearly in Fig.~\ref{fig:scale}(a).
The time scale $t_1$ coincides with the beginning of the equilibration of $P(2n)$, i.e., the detailed balancing of the creation and annihilation of quasiparticles.
The metastable period from $t_1$ to $t_2$ is characterized by the breaking of the spatial inversion symmetry, i.e., $C_{\bar 0\bar 1}(d)\not=C_{\bar1\bar0}(d)$, along with the invariance of $C_{\bar x\bar y}(d)$ in time.
From $t\simeq t_2$, the imbalance between $C_{\bar 0\bar 1}(d)$ and $C_{\bar1\bar0}(d)$ begins to relax and the entanglement entropy grows logarithmically in time.
The final equilibrium is reached at $t\simeq t_3$, which is signaled by $C_{\bar 0\bar 1}(d)=C_{\bar1\bar0}(d)$.

\begin{figure}
	\begin{center}
		\includegraphics[width=0.48\textwidth]{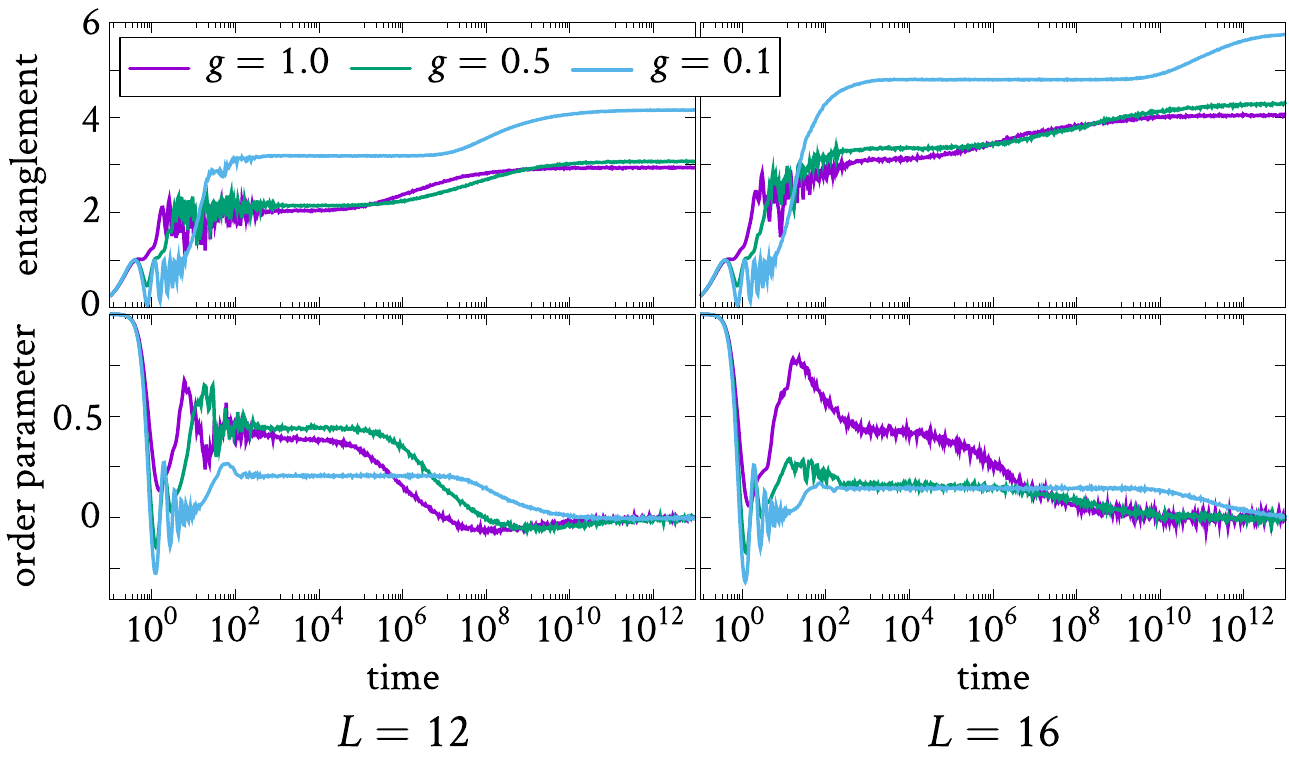}
	\end{center}
	\vspace{-1em}
	\caption{Average time evolution of the half-chain entanglement and the order parameter $\mathcal{C}$ for $D=0.1$ and $J=1$.
	}
	\label{fig:metasta}
\end{figure}

\begin{figure}[b]
\begin{center}
\includegraphics[width=0.46\textwidth]{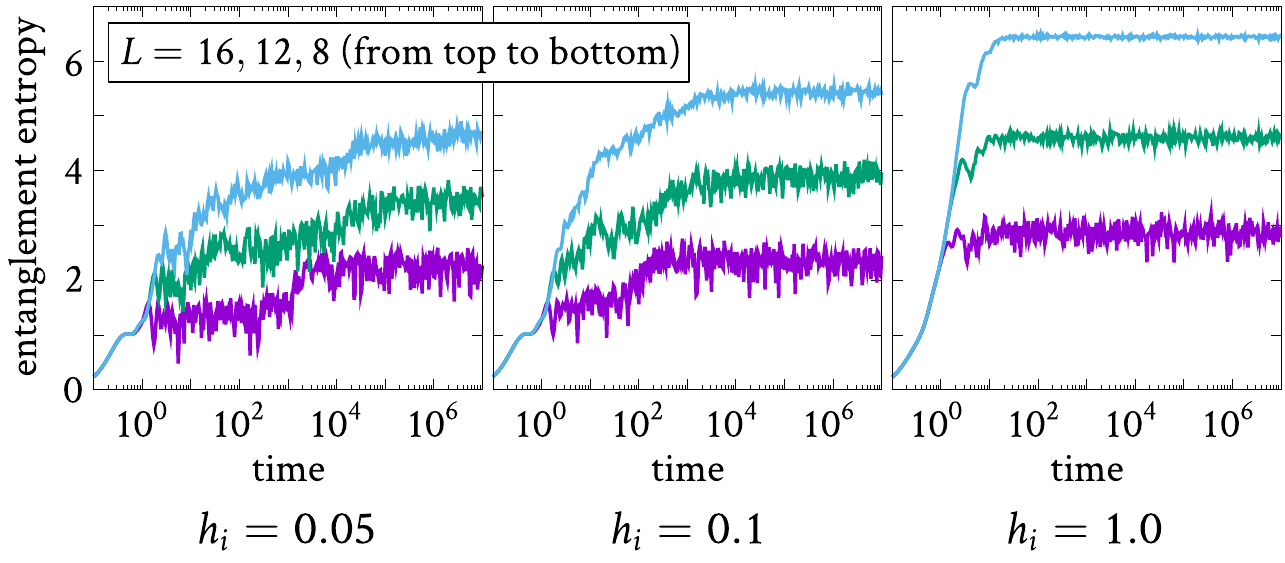}
\end{center}
\vspace{-1em}
\caption{Time evolution of the half-chain entanglement entropy for constant $h_i$ with $J=g=1$.}
\label{fig:disorder-free}
\end{figure}

Summing up, $t_1$, $t_2$, and $t_3$ are signaled, respectively, by the equilibration of $P(2n)$, the relaxation of $C_{\bar 0\bar 1}(d)\not=C_{\bar1\bar0}(d)$, and $C_{\bar 0\bar 1}(d)=C_{\bar1\bar0}(d)$.
By defining the time scales in this way, one can interpret the transition in Fig.~\ref{fig:scale}(b) as follows.
For sufficiently small $D$, the time scales are clearly separated as $t_1< t_2< t_3$.
As $D$ increases, all the time scales decrease, but $t_1$ decreases much more slowly than the others.
As a result, the duration of the first plateau decreases and for $D\simeq0.5$, the time scales reach $t_1\simeq t_2<t_3$ and thus the entanglement dynamics resembles that of the conventional MBL in one dimension.
As $D$ increases further, the time scales enter the regime of $t_2<t_1<t_3$ and the signature of the $\log(t)$ behavior gradually fades away until the discrimination of the time scales becomes meaningless.


\textit{Metastability and disorder-free MBL.---}%
The metastable period is largely elongated by decreasing $g$.
In that case, the first term in the Hamiltonian~\eqref{eq:quasi} is dominant and again the situation is analogous to the case of prethermalization~\cite{ber15,mor18}.
Fig.~\ref{fig:metasta} depicts the evolution of the entanglement entropy and the order parameter defined as
\begin{equation}
	\mathcal{C} = \sum_{d=1}^{L/4} \ll
		\mathcal{C}_{\bar0\bar1}(d)-\mathcal{C}_{\bar1\bar0}(d).
	\rr
\end{equation}
For example, for $D=g=0.1$, the relaxation dynamics becomes very slow, and the signatures of metastability and MBL coexist.

Another interesting feature is that the MBL signatures are retained in the absence of disorder.
Fig.~\ref{fig:disorder-free} depicts the evolution of the entanglement entropy when $h_i$ is a site-independent constant.
As was predicted above, the figure clearly shows a $\log(t)$ scaling behavior for small $h_i$.
In addition, the metastability is again observed simultaneously.

\begin{figure}[t]
\begin{center}
\includegraphics[width=0.46\textwidth]{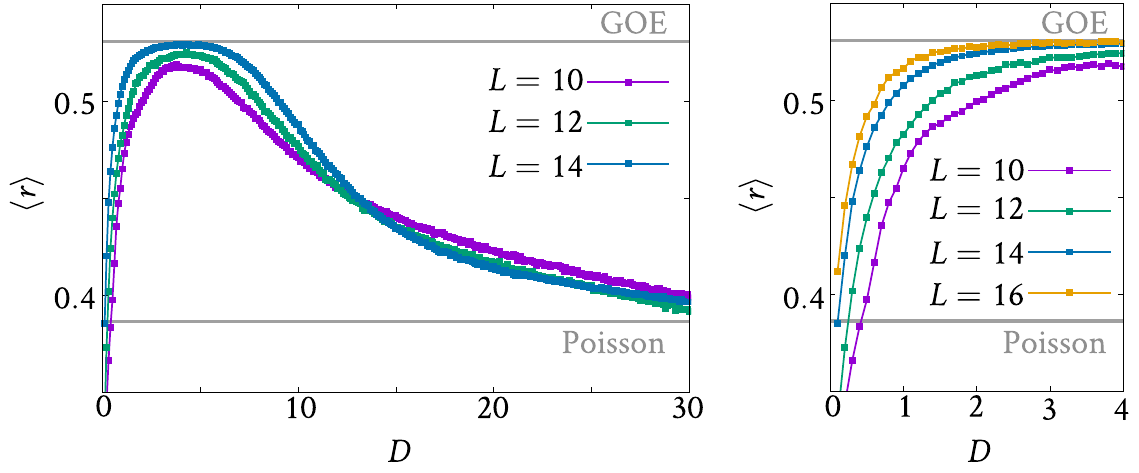}
\end{center}
\vspace{-1em}
\caption{Average of the level-spacing data $r_n$ with respect to $D$ for $J=g=1$ and the periodic boundary condition.
The right plot is a zoom on the small $D$ part of the left one.
{The GOE average is taken from Ref.~\cite{ata13}.}
}
\label{fig:r}
\end{figure}

\begin{figure}[b]
\begin{center}
\includegraphics[width=0.46\textwidth]{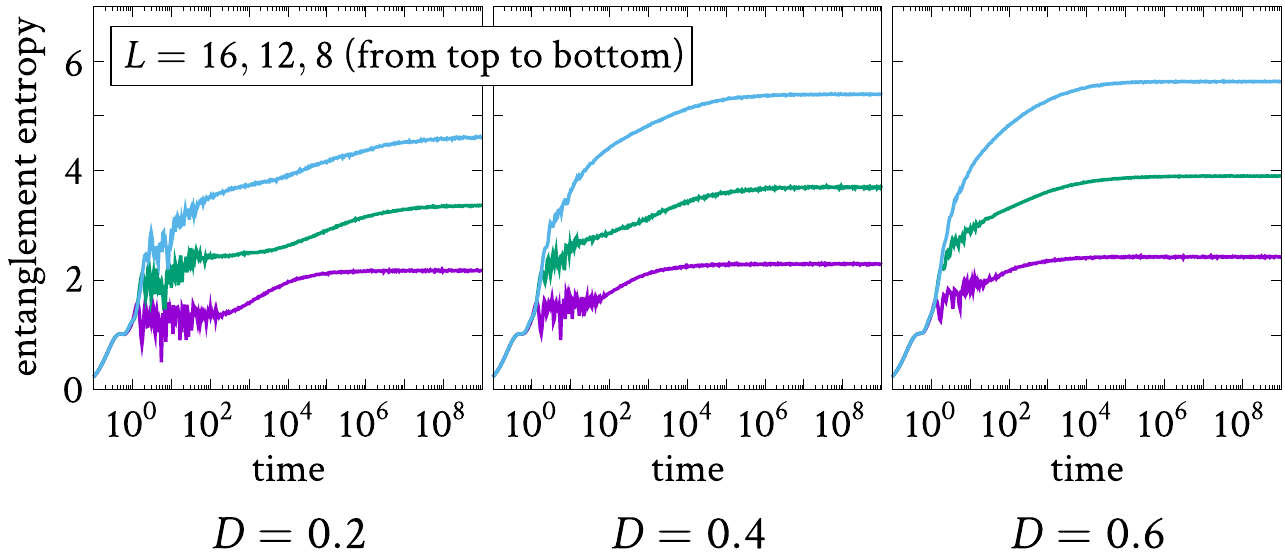}
\end{center}
\vspace{-1em}
\caption{Time evolution of the half-chain entanglement entropy for $J=g=1$.
}
\label{fig:thermo}
\end{figure}


\textit{Thermodynamic limit.---}%
An important question is whether the same signatures for small $D$ are retained for large $L$.
We speculate the following in view of our numerical results.
First, the weak-to-intermediate-disorder transition appears to be a crossover.
To see this, we plot in Fig.~\ref{fig:r} the average of $r_n$ with respect to $D$ for different $L$, as was done in Ref.~\cite{oga07}.
Here we have taken the periodic boundary condition to compare the values with those of the Poissonian and GOE.
It is observed that the three curves cross around $D\sim13$, which indicates that the intermediate-to-strong-disorder transition is indeed a phase transition~\cite{pal10,hus13,kja14,nan15}.
However, a similar feature is not observed for small $D$.
The next question is then what happens to the parameter window for small $D$, which enables the observation of the slow relaxation dynamics.
For this, we obtained Fig.~\ref{fig:thermo}, which compares the entanglement dynamics for different $L$ and $D$.
In view of Figs.~\ref{fig:metasta}--\ref{fig:thermo}, there appears to be a trade-off:
when $L$ increases, while the relaxation time is elongated, the system enters the intermediate regime more quickly by increasing $D$, which means the parameter window for the slow dynamics becomes narrower.


\textit{Discussion.---}%
In conclusion, we have demonstrated a system exhibiting mixed signatures of disordered and disorder-free MBL, and prethermalization, and ascribed them to the quasiparticle dynamics.
We suspect that the crosstalk between different subspaces $\mathcal{H}_{2n}$ hinders a clear-cut MBL phase for small $D$ when $L$ is large.
One could instead consider Hamiltonian~\eqref{eq:quasi} from the beginning and adjust transition terms appropriately to produce different situations, which we leave as future work.
In a slightly different context, the proposed effective disorder mechanism may offer a new conceptual ground of non-ergodic many-body dynamics on top of those under active exploration, such as many-body scars and fracton dynamics~\cite{shi17,tur18,pai19,sal19,khe19}, as well as those mentioned in the introduction.
In this context, another distinct feature of our mechanism is that the interaction is of translationally invariant two-body nearest-neighbor, which would be more relevant to the experimental demonstration.


\begin{acknowledgments}
We thank A. Lakshminarayan, C.-Y. Park, and A. Andreanov for useful discussions. 
YZ acknowledges support by Fundamental Research Funds for the Central Universities (3102017OQD074).
JC acknowledges support by the R\&D Convergence Program of NST (National Research
Council of Science and Technology), the Ministry of Science, ICT \& Future Planning, Gyeongsangbuk-do and Pohang City (Grant No.  CAP-15-08-KRISS).
\end{acknowledgments}

\section{Appendix}

\subsection{Duality Transformation}

\setcounter{equation}{0}
\setcounter{figure}{0}
\renewcommand{\theequation}{A\arabic{equation}}
\renewcommand{\thefigure}{A\arabic{figure}}

Consider the Hamiltonian
\begin{equation}
H = \sum_{i=1}^{L/2} 
\ll \sum_{\alpha=1}^2 
\(
h_i \sigma_{i,\alpha}^x
- J \sigma_{i,\alpha}^z \sigma_{i+1,\alpha}^z
\)
- g \sigma_{i,1}^x \sigma_{i,2}^x
\rr
\end{equation}
with $\sigma^z_{L/2+1}=0$.
Let us first apply to the $\alpha=2$ chain the duality transformation
\begin{equation}
\begin{split}
	&\sigma^z \(i+\tfrac12\) = \prod_{k=1}^i \sigma_{k,2}^x,\\
	&\sigma^x \(i+\tfrac12\) = \sigma_{i,2}^z \sigma_{i+1,2}^z,
\end{split}
\end{equation}
which implies
\begin{equation}
\begin{split}
	&\sigma_{i,2}^x 
	= \sigma^z \(i-\tfrac12\) \sigma^z \(i+\tfrac12\),\\
	&\sigma^z_{i,2}\sigma^z_{i+1,2} 
	= \sigma^x \(i+\tfrac12\),
\end{split}
\end{equation}
where it is understood that $\sigma^z\(\tfrac12\)=1$.
Letting $\sigma^x (i)=\sigma^x_{i,1}$ and $\sigma^z (i)=\sigma^z_{i,1}$, it then follows, omitting $\sum_{i=1}^{L/2}$, that
\begin{equation}
\begin{split}
	H &= h_i \ll
	\sigma^x(i)
	+\sigma^z\(i-\tfrac12\)\sigma^z\(i+\tfrac12\)
	\rr \\
	&\quad -J \ll
	\sigma^z(i)\sigma^z(i+1)+\sigma^x\(i+\tfrac12\)
	\rr \\
	&\quad -g \sigma^x(i)\sigma^z\(i-\tfrac12\)
	\sigma^z\(i+\tfrac12\) \\
	&= h_i \ll
	\sigma^x(i) 
	+\sigma^z\(i-\tfrac12\)\sigma^z(i)\sigma^z(i)
	\sigma^z\(i+\tfrac12\)
	\rr \\
	&\quad -J \ll
	\sigma^z(i)\sigma^z\(i+\tfrac12\)
	\sigma^z\(i+\tfrac12\)\sigma^z(i+1) \right. \\
	&\quad\quad\quad  \left. +\sigma^x\(i+\tfrac12\)
	\rr \\
	&\quad -g \sigma^x(i)\sigma^z\(i-\tfrac12\)
	\sigma^z(i)\sigma^z(i)\sigma^z\(i+\tfrac12\).
\end{split}
\end{equation}
Applying another duality transformation
\begin{equation}
\begin{split}
	&\sigma^x(i) \rightarrow 
	\sigma^z\(i-\tfrac14\)\sigma^z\(i+\tfrac14\),\\
	&\sigma^z(i)\sigma^z\(i+\tfrac12\) \rightarrow
	\sigma^x\(i+\tfrac14\),
\end{split}
\end{equation}
we have
\begin{equation}
\begin{split}
	H &= h_i \ll
	\sigma^z\(i-\tfrac14\)\sigma^z\(i+\tfrac14\)
	-\sigma^x\(i-\tfrac14\)\sigma^x\(i+\tfrac14\)
	\rr \\
	&\quad -J \ll
	\sigma^x\(i+\tfrac14\)\sigma^x\(i+\tfrac34\)
	-\sigma^z\(i+\tfrac14\)\sigma^z\(i+\tfrac34\)
	\rr \\
	&\quad +g\sigma^y\(i-\tfrac14\)\sigma^y\(i+\tfrac14\).
\end{split}
\end{equation}
Simply by reindexing and applying a unitary transformation
\begin{equation}
\sigma^x\rightarrow\sigma^z,\quad
\sigma^z\rightarrow\sigma^x,\quad
\sigma^y\rightarrow-\sigma^y,
\end{equation}
we obtain the final Hamiltonian
\begin{equation}
\begin{split}
H &= \sum_{i=1}^{L/2} \ll
h_i \(
\sigma_{2i-1}^x \sigma_{2i}^x
+ \sigma_{2i-1}^y \sigma_{2i}^y
\)
+ g \sigma_{2i-1}^z \sigma_{2i}^z \right. \\
&\quad\quad\quad \left. - J \(
\sigma_{2i}^x \sigma_{2i+1}^x
+ \sigma_{2i}^y \sigma_{2i+1}^y
\)
\rr.
\end{split}
\end{equation}

\subsection{Emergence of effective disorder}

\begin{figure*}
\includegraphics[width=0.99\textwidth]{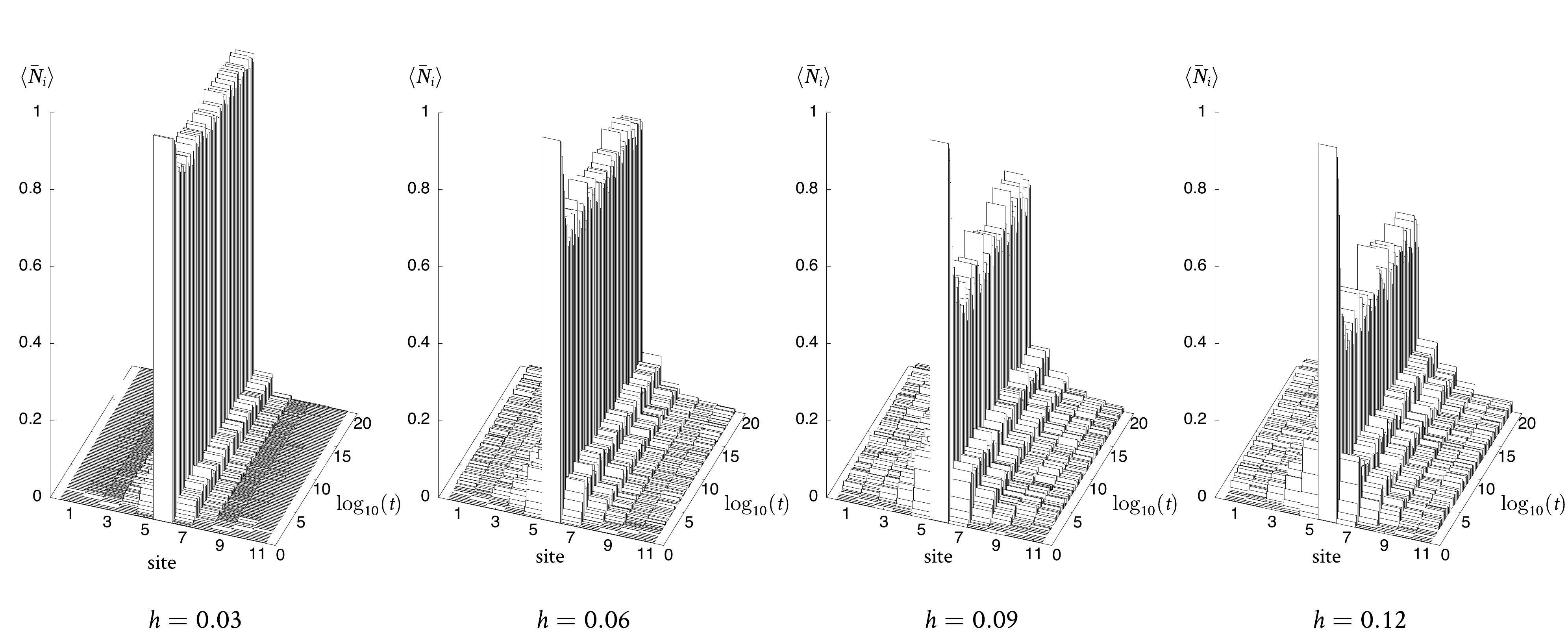}
\caption{Time evolution of the quasiparticle distribution over site $i$ for $J=g=1$. $\langle{\bar N_i}\rangle$ denotes $\bra{\Psi(t)}\bar N_i\ket{\Psi(t)}$ at time $t$. Here, $L=11$ and the initial state is $\ket{\Psi(0)}=\ket{\bar q}_6\prod_{i\not=6}\ket 0_i$. Note that the time axis is in logarithmic scale.}
\label{fig:1}
\end{figure*}

This section is aimed at testing the validity of our mechanism of emergent disorder.
To do so, we consider a tailored model that makes the signature of localization more explicit while retaining the essence of the quasiparticle picture.
Specifically, we consider spins of dimension three having a single quasiparticle state $\ket{\bar q}$, instead of $\ket{\bar 0}$ and $\ket{\bar 1}$, in addition to the vacuum states $\ket 0$ and $\ket 1$.
Moreover, we keep the number of quasiparticles to be one, i.e., $\sum_i \ket{\bar q}_i\bra{\bar q}=1$.
This model exhibits Anderson localization without any inherent disorder in the Hamiltonian.

As in the main text, we define the Pauli operators for $\{\ket0,\ket1\}$ as
\begin{equation}
	X_i=\ket0_i\bra1+\ket1_i\bra0,\quad Z_i=\ket0_i\bra0-\ket1_i\bra1,
\end{equation}
and the number operator for the quasiparticle as
\begin{equation}
	\bar N_i=\ket{\bar q}_i\bra{\bar q}.
\end{equation}
As mentioned, we will keep $\sum_i \bar N_i=1$.
Consider Hamiltonian
\begin{equation}
	H=H_0+H_1,
\end{equation}
where
\begin{equation}
	\begin{split}
	H_0 &= \sum_{i=1}^{L-1} (-2JX_i -gZ_iZ_{i+1}-gZ_i \bar N_{i+1}+g\bar N_iZ_{i+1}),\\
	H_1 &= -h\sum_{i=1}^{L-1} (\ket0_i\bra{\bar q} \otimes \ket{\bar q}_{i+1}\bra1+
	\ket1_i\bra{\bar q} \otimes \ket{\bar q}_{i+1}\bra0 \\
	&\quad\quad\quad\quad\quad + \text{H.c.}).
	\end{split}
	\label{eq:hamil}
\end{equation}
$H_0$ resembles Hamiltonian (7) in the main text except that the last term is absent here because $\sum_i \bar N_i=1$.
$H_1$ models the second line in Eq. (8) with $h$ a {\em constant} hopping amplitude.
There is no creation or annihilation of quasiparticles here.
%
Fig.~\ref{fig:1} shows the dynamics of a single quasiparticle for different $h$. 
Here, $L=11$ and the initial state is $\ket{\Psi(0)}=\ket{\bar q}_6\prod_{i\not=6}\ket 0_i$.
It is clearly seen that the system is Anderson-localised.
Note that the time axis is in logarithmic scale.
Here, smaller $h$ corresponds to stronger effective disorder because the energy scale of $H_0$ in Eq.~\eqref{eq:hamil} determines the strength of effective disorder.

\begin{figure*}
	\includegraphics[width=0.99\textwidth]{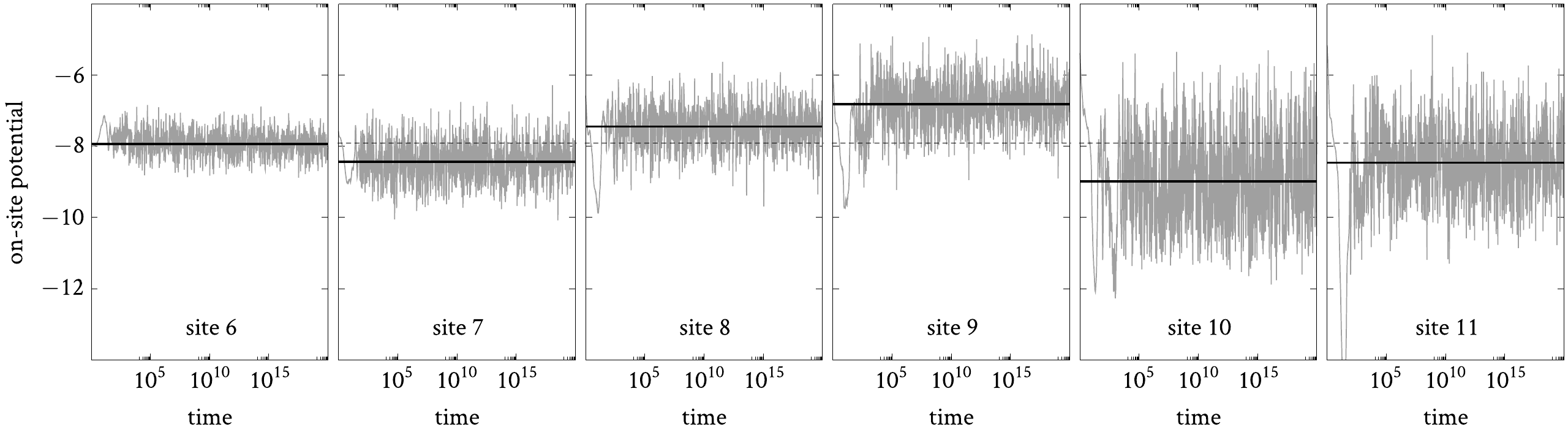}
	\caption{The on-site potential acting on the quasiparticle in a rotating frame for $J=g=1$ and $h=0.1$. The solid line indicates the average potential after the equilibration (averaged over $10^5\le t\le 10^{20}$). The dotted line indicates the average potential at the central site 6, which is added for comparison.}
	\label{fig:2}
\end{figure*}

At first, this result seems to contradict our knowledge that dynamic disorder breaks Anderson localization~\cite{mad77}.
In order to resolve this issue, let us obtain the explicit expression for the effective quasiparticle Hamiltonian.
We write the state in a rotating frame as
\begin{equation}
    \ket{\Psi(t)} = \sum_{j=1}^L c_j(t)e^{iR_j(t)}\ket{\bar q}_j \ket{\phi_j(t)},
\end{equation}
where $\ket{\phi_j(t)}$ denotes the associated (normalized) vacuum state spanned by $\{\ket0_k,\ket1_k\}$ with $k\not=j$.
Choosing 
\begin{equation}
    R_j(t)=i\int_0^t dt' \bigl<\phi_j(t')\bigr|\bigl.\dot\phi_j(t')\bigr>,
\end{equation}
the time evolution is given by
\begin{equation}
    i\dot c_j(t) = \sum_{k=1}^L M_{jk}(t)c_k(t),
\end{equation}
where
\begin{equation}
    M_{jk}(t) = 
    \begin{cases}
    {}_j\bra{\bar q} \bra{\phi_j(t)}H_0\ket{\bar q}_j\ket{\phi_j(t)} & j=k,\\
    e^{-iR_{jk}(t)} {}_j\bra{\bar q}\bra{\phi_j(t)}H_1\ket{\bar q}_k\ket{\phi_k(t)} & j\not=k.
    \end{cases}
\end{equation}
Here, $R_{jk}(t)=R_j(t)-R_k(t)$.
Now we can define the quasiparticle Hamiltonian as
\begin{equation}
\bar H(t)=\sum_{j,k=1}^L M_{jk}(t)\ket{\bar q}_j{}_k\bra{\bar q}.
\end{equation}
Then, the quasiparticle state $\ket{\bar\Psi(t)}=\sum_j c_j(t)\ket{\bar q}_j$ is governed by
\begin{equation}
i\ket{\dot{\bar\Psi}(t)}=\bar H(t)\ket{\bar{\Psi}(t)}.
\end{equation}

Fig.~\ref{fig:2} shows the diagonal elements of $\bar H(t)$ for $J=g=1$ and $h=0.1$.
As can be seen, the quasiparticle is subjected to dynamic on-site disorder.
However, the disorder here is different in nature from conventionally studied ones such as Gaussian white noise~\cite{mad77}.
A notable feature is that the disorder potential fluctuates in time around a seemingly random, nonzero average.
This implies that the effective disorder in our system is interpreted as having an infinite correlation time, which can localize the system despite being time-dependent.

\bibliography{reference}

\begin{thebibliography}{37}%
\makeatletter
\providecommand \@ifxundefined [1]{%
 \@ifx{#1\undefined}
}%
\providecommand \@ifnum [1]{%
 \ifnum #1\expandafter \@firstoftwo
 \else \expandafter \@secondoftwo
 \fi
}%
\providecommand \@ifx [1]{%
 \ifx #1\expandafter \@firstoftwo
 \else \expandafter \@secondoftwo
 \fi
}%
\providecommand \natexlab [1]{#1}%
\providecommand \enquote  [1]{``#1''}%
\providecommand \bibnamefont  [1]{#1}%
\providecommand \bibfnamefont [1]{#1}%
\providecommand \citenamefont [1]{#1}%
\providecommand \href@noop [0]{\@secondoftwo}%
\providecommand \href [0]{\begingroup \@sanitize@url \@href}%
\providecommand \@href[1]{\@@startlink{#1}\@@href}%
\providecommand \@@href[1]{\endgroup#1\@@endlink}%
\providecommand \@sanitize@url [0]{\catcode `\\12\catcode `\$12\catcode
  `\&12\catcode `\#12\catcode `\^12\catcode `\_12\catcode `\%12\relax}%
\providecommand \@@startlink[1]{}%
\providecommand \@@endlink[0]{}%
\providecommand \url  [0]{\begingroup\@sanitize@url \@url }%
\providecommand \@url [1]{\endgroup\@href {#1}{\urlprefix }}%
\providecommand \urlprefix  [0]{URL }%
\providecommand \Eprint [0]{\href }%
\providecommand \doibase [0]{http://dx.doi.org/}%
\providecommand \selectlanguage [0]{\@gobble}%
\providecommand \bibinfo  [0]{\@secondoftwo}%
\providecommand \bibfield  [0]{\@secondoftwo}%
\providecommand \translation [1]{[#1]}%
\providecommand \BibitemOpen [0]{}%
\providecommand \bibitemStop [0]{}%
\providecommand \bibitemNoStop [0]{.\EOS\space}%
\providecommand \EOS [0]{\spacefactor3000\relax}%
\providecommand \BibitemShut  [1]{\csname bibitem#1\endcsname}%
\let\auto@bib@innerbib\@empty
\bibitem [{\citenamefont {Jaynes}(1957)}]{jay57}%
  \BibitemOpen
  \bibfield  {author} {\bibinfo {author} {\bibfnamefont {E.~T.}\ \bibnamefont
  {Jaynes}},\ }\bibfield  {title} {\enquote {\bibinfo {title} {Information
  theory and statistical mechanics},}\ }\href {\doibase
  10.1103/PhysRev.106.620} {\bibfield  {journal} {\bibinfo  {journal} {Phys.
  Rev.}\ }\textbf {\bibinfo {volume} {106}},\ \bibinfo {pages} {620} (\bibinfo
  {year} {1957})}\BibitemShut {NoStop}%
\bibitem [{\citenamefont {Rigol}\ \emph {et~al.}(2007)\citenamefont {Rigol},
  \citenamefont {Dunjko}, \citenamefont {Yurovsky},\ and\ \citenamefont
  {Olshanii}}]{rig07}%
  \BibitemOpen
  \bibfield  {author} {\bibinfo {author} {\bibfnamefont {M.}~\bibnamefont
  {Rigol}}, \bibinfo {author} {\bibfnamefont {V.}~\bibnamefont {Dunjko}},
  \bibinfo {author} {\bibfnamefont {V.}~\bibnamefont {Yurovsky}}, \ and\
  \bibinfo {author} {\bibfnamefont {M.}~\bibnamefont {Olshanii}},\ }\bibfield
  {title} {\enquote {\bibinfo {title} {Relaxation in a completely integrable
  many-body quantum system: An ab initio study of the dynamics of the highly
  excited states of 1d lattice hard-core bosons},}\ }\href {\doibase
  10.1103/PhysRevLett.98.050405} {\bibfield  {journal} {\bibinfo  {journal}
  {Phys. Rev. Lett.}\ }\textbf {\bibinfo {volume} {98}},\ \bibinfo {pages}
  {050405} (\bibinfo {year} {2007})}\BibitemShut {NoStop}%
\bibitem [{\citenamefont {Mori}\ \emph {et~al.}(2018)\citenamefont {Mori},
  \citenamefont {Ikeda}, \citenamefont {Kaminishi},\ and\ \citenamefont
  {Ueda}}]{mor18}%
  \BibitemOpen
  \bibfield  {author} {\bibinfo {author} {\bibfnamefont {T.}~\bibnamefont
  {Mori}}, \bibinfo {author} {\bibfnamefont {T.~N.}\ \bibnamefont {Ikeda}},
  \bibinfo {author} {\bibfnamefont {E.}~\bibnamefont {Kaminishi}}, \ and\
  \bibinfo {author} {\bibfnamefont {M.}~\bibnamefont {Ueda}},\ }\bibfield
  {title} {\enquote {\bibinfo {title} {Thermalization and prethermalization in
  isolated quantum systems: a theoretical overview},}\ }\href {\doibase
  10.1088/1361-6455/aabcdf} {\bibfield  {journal} {\bibinfo  {journal} {J.
  Phys. B}\ }\textbf {\bibinfo {volume} {51}},\ \bibinfo {pages} {112001}
  (\bibinfo {year} {2018})}\BibitemShut {NoStop}%
\bibitem [{\citenamefont {Basko}\ \emph {et~al.}(2006)\citenamefont {Basko},
  \citenamefont {Aleiner},\ and\ \citenamefont {Altshuler}}]{bas06}%
  \BibitemOpen
  \bibfield  {author} {\bibinfo {author} {\bibfnamefont {D.~M.}\ \bibnamefont
  {Basko}}, \bibinfo {author} {\bibfnamefont {I.~L.}\ \bibnamefont {Aleiner}},
  \ and\ \bibinfo {author} {\bibfnamefont {B.~L.}\ \bibnamefont {Altshuler}},\
  }\bibfield  {title} {\enquote {\bibinfo {title} {Metal–insulator transition
  in a weakly interacting many-electron system with localized single-particle
  states},}\ }\href {\doibase https://doi.org/10.1016/j.aop.2005.11.014}
  {\bibfield  {journal} {\bibinfo  {journal} {Ann. Phys. (N. Y.)}\ }\textbf
  {\bibinfo {volume} {321}},\ \bibinfo {pages} {1126} (\bibinfo {year}
  {2006})}\BibitemShut {NoStop}%
\bibitem [{\citenamefont {Oganesyan}\ and\ \citenamefont {Huse}(2007)}]{oga07}%
  \BibitemOpen
  \bibfield  {author} {\bibinfo {author} {\bibfnamefont {V.}~\bibnamefont
  {Oganesyan}}\ and\ \bibinfo {author} {\bibfnamefont {D.~A.}\ \bibnamefont
  {Huse}},\ }\bibfield  {title} {\enquote {\bibinfo {title} {Localization of
  interacting fermions at high temperature},}\ }\href {\doibase
  10.1103/PhysRevB.75.155111} {\bibfield  {journal} {\bibinfo  {journal} {Phys.
  Rev. B}\ }\textbf {\bibinfo {volume} {75}},\ \bibinfo {pages} {155111}
  (\bibinfo {year} {2007})}\BibitemShut {NoStop}%
\bibitem [{\citenamefont {\v{Z}nidari\v{c}}\ \emph {et~al.}(2008)\citenamefont
  {\v{Z}nidari\v{c}}, \citenamefont {Prosen},\ and\ \citenamefont
  {Prelov\v{s}ek}}]{zni08}%
  \BibitemOpen
  \bibfield  {author} {\bibinfo {author} {\bibfnamefont {M.}~\bibnamefont
  {\v{Z}nidari\v{c}}}, \bibinfo {author} {\bibfnamefont {T.}~\bibnamefont
  {Prosen}}, \ and\ \bibinfo {author} {\bibfnamefont {P.}~\bibnamefont
  {Prelov\v{s}ek}},\ }\bibfield  {title} {\enquote {\bibinfo {title} {Many-body
  localization in the heisenberg ${XXZ}$ magnet in a random field},}\ }\href
  {\doibase 10.1103/PhysRevB.77.064426} {\bibfield  {journal} {\bibinfo
  {journal} {Phys. Rev. B}\ }\textbf {\bibinfo {volume} {77}},\ \bibinfo
  {pages} {064426} (\bibinfo {year} {2008})}\BibitemShut {NoStop}%
\bibitem [{\citenamefont {Pal}\ and\ \citenamefont {Huse}(2010)}]{pal10}%
  \BibitemOpen
  \bibfield  {author} {\bibinfo {author} {\bibfnamefont {A.}~\bibnamefont
  {Pal}}\ and\ \bibinfo {author} {\bibfnamefont {D.~A.}\ \bibnamefont {Huse}},\
  }\bibfield  {title} {\enquote {\bibinfo {title} {Many-body localization phase
  transition},}\ }\href {\doibase 10.1103/PhysRevB.82.174411} {\bibfield
  {journal} {\bibinfo  {journal} {Phys. Rev. B}\ }\textbf {\bibinfo {volume}
  {82}},\ \bibinfo {pages} {174411} (\bibinfo {year} {2010})}\BibitemShut
  {NoStop}%
\bibitem [{\citenamefont {Huse}\ \emph {et~al.}(2013)\citenamefont {Huse},
  \citenamefont {Nandkishore}, \citenamefont {Oganesyan}, \citenamefont {Pal},\
  and\ \citenamefont {Sondhi}}]{hus13}%
  \BibitemOpen
  \bibfield  {author} {\bibinfo {author} {\bibfnamefont {D.~A.}\ \bibnamefont
  {Huse}}, \bibinfo {author} {\bibfnamefont {R.}~\bibnamefont {Nandkishore}},
  \bibinfo {author} {\bibfnamefont {V.}~\bibnamefont {Oganesyan}}, \bibinfo
  {author} {\bibfnamefont {A.}~\bibnamefont {Pal}}, \ and\ \bibinfo {author}
  {\bibfnamefont {S.~L.}\ \bibnamefont {Sondhi}},\ }\bibfield  {title}
  {\enquote {\bibinfo {title} {Localization-protected quantum order},}\ }\href
  {\doibase 10.1103/PhysRevB.88.014206} {\bibfield  {journal} {\bibinfo
  {journal} {Phys. Rev. B}\ }\textbf {\bibinfo {volume} {88}},\ \bibinfo
  {pages} {014206} (\bibinfo {year} {2013})}\BibitemShut {NoStop}%
\bibitem [{\citenamefont {Kj\"all}\ \emph {et~al.}(2014)\citenamefont
  {Kj\"all}, \citenamefont {Bardarson},\ and\ \citenamefont
  {Pollmann}}]{kja14}%
  \BibitemOpen
  \bibfield  {author} {\bibinfo {author} {\bibfnamefont {J.~A.}\ \bibnamefont
  {Kj\"all}}, \bibinfo {author} {\bibfnamefont {J.~H.}\ \bibnamefont
  {Bardarson}}, \ and\ \bibinfo {author} {\bibfnamefont {F.}~\bibnamefont
  {Pollmann}},\ }\bibfield  {title} {\enquote {\bibinfo {title} {Many-body
  localization in a disordered quantum ising chain},}\ }\href {\doibase
  10.1103/PhysRevLett.113.107204} {\bibfield  {journal} {\bibinfo  {journal}
  {Phys. Rev. Lett.}\ }\textbf {\bibinfo {volume} {113}},\ \bibinfo {pages}
  {107204} (\bibinfo {year} {2014})}\BibitemShut {NoStop}%
\bibitem [{\citenamefont {Nandkishore}\ and\ \citenamefont
  {Huse}(2015)}]{nan15}%
  \BibitemOpen
  \bibfield  {author} {\bibinfo {author} {\bibfnamefont {R.}~\bibnamefont
  {Nandkishore}}\ and\ \bibinfo {author} {\bibfnamefont {D.~A.}\ \bibnamefont
  {Huse}},\ }\bibfield  {title} {\enquote {\bibinfo {title} {Many-body
  localization and thermalization in quantum statistical mechanics},}\ }\href
  {\doibase 10.1146/annurev-conmatphys-031214-014726} {\bibfield  {journal}
  {\bibinfo  {journal} {Annu. Rev. Condens. Matter Phys.}\ }\textbf {\bibinfo
  {volume} {6}},\ \bibinfo {pages} {15--38} (\bibinfo {year}
  {2015})}\BibitemShut {NoStop}%
\bibitem [{\citenamefont {Bardarson}\ \emph {et~al.}(2012)\citenamefont
  {Bardarson}, \citenamefont {Pollmann},\ and\ \citenamefont {Moore}}]{bar12}%
  \BibitemOpen
  \bibfield  {author} {\bibinfo {author} {\bibfnamefont {J.~H.}\ \bibnamefont
  {Bardarson}}, \bibinfo {author} {\bibfnamefont {F.}~\bibnamefont {Pollmann}},
  \ and\ \bibinfo {author} {\bibfnamefont {J.~E.}\ \bibnamefont {Moore}},\
  }\bibfield  {title} {\enquote {\bibinfo {title} {Unbounded growth of
  entanglement in models of many-body localization},}\ }\href {\doibase
  10.1103/PhysRevLett.109.017202} {\bibfield  {journal} {\bibinfo  {journal}
  {Phys. Rev. Lett.}\ }\textbf {\bibinfo {volume} {109}},\ \bibinfo {pages}
  {017202} (\bibinfo {year} {2012})}\BibitemShut {NoStop}%
\bibitem [{\citenamefont {Vosk}\ and\ \citenamefont {Altman}(2013)}]{vos13}%
  \BibitemOpen
  \bibfield  {author} {\bibinfo {author} {\bibfnamefont {R.}~\bibnamefont
  {Vosk}}\ and\ \bibinfo {author} {\bibfnamefont {E.}~\bibnamefont {Altman}},\
  }\bibfield  {title} {\enquote {\bibinfo {title} {Many-body localization in
  one dimension as a dynamical renormalization group fixed point},}\ }\href
  {\doibase 10.1103/PhysRevLett.110.067204} {\bibfield  {journal} {\bibinfo
  {journal} {Phys. Rev. Lett.}\ }\textbf {\bibinfo {volume} {110}},\ \bibinfo
  {pages} {067204} (\bibinfo {year} {2013})}\BibitemShut {NoStop}%
\bibitem [{\citenamefont {Andraschko}\ \emph {et~al.}(2014)\citenamefont
  {Andraschko}, \citenamefont {Enss},\ and\ \citenamefont {Sirker}}]{and14}%
  \BibitemOpen
  \bibfield  {author} {\bibinfo {author} {\bibfnamefont {F.}~\bibnamefont
  {Andraschko}}, \bibinfo {author} {\bibfnamefont {T.}~\bibnamefont {Enss}}, \
  and\ \bibinfo {author} {\bibfnamefont {J.}~\bibnamefont {Sirker}},\
  }\bibfield  {title} {\enquote {\bibinfo {title} {Purification and many-body
  localization in cold atomic gases},}\ }\href {\doibase
  10.1103/PhysRevLett.113.217201} {\bibfield  {journal} {\bibinfo  {journal}
  {Phys. Rev. Lett.}\ }\textbf {\bibinfo {volume} {113}},\ \bibinfo {pages}
  {217201} (\bibinfo {year} {2014})}\BibitemShut {NoStop}%
\bibitem [{\citenamefont {Imbrie}(2016{\natexlab{a}})}]{imb16}%
  \BibitemOpen
  \bibfield  {author} {\bibinfo {author} {\bibfnamefont {J.~Z.}\ \bibnamefont
  {Imbrie}},\ }\bibfield  {title} {\enquote {\bibinfo {title} {Diagonalization
  and many-body localization for a disordered quantum spin chain},}\ }\href
  {\doibase 10.1103/PhysRevLett.117.027201} {\bibfield  {journal} {\bibinfo
  {journal} {Phys. Rev. Lett.}\ }\textbf {\bibinfo {volume} {117}},\ \bibinfo
  {pages} {027201} (\bibinfo {year} {2016}{\natexlab{a}})}\BibitemShut
  {NoStop}%
\bibitem [{\citenamefont {Imbrie}(2016{\natexlab{b}})}]{imb16b}%
  \BibitemOpen
  \bibfield  {author} {\bibinfo {author} {\bibfnamefont {J.~Z.}\ \bibnamefont
  {Imbrie}},\ }\bibfield  {title} {\enquote {\bibinfo {title} {On many-body
  localization for quantum spin chains},}\ }\href {\doibase
  10.1007/s10955-016-1508-x} {\bibfield  {journal} {\bibinfo  {journal} {J.
  Stat. Phys.}\ }\textbf {\bibinfo {volume} {163}},\ \bibinfo {pages} {998}
  (\bibinfo {year} {2016}{\natexlab{b}})}\BibitemShut {NoStop}%
\bibitem [{\citenamefont {Schreiber}\ \emph {et~al.}(2015)\citenamefont
  {Schreiber}, \citenamefont {Hodgman}, \citenamefont {Bordia}, \citenamefont
  {L{\"u}schen}, \citenamefont {Fischer}, \citenamefont {Vosk}, \citenamefont
  {Altman}, \citenamefont {Schneider},\ and\ \citenamefont {Bloch}}]{sch15}%
  \BibitemOpen
  \bibfield  {author} {\bibinfo {author} {\bibfnamefont {M.}~\bibnamefont
  {Schreiber}}, \bibinfo {author} {\bibfnamefont {S.~S.}\ \bibnamefont
  {Hodgman}}, \bibinfo {author} {\bibfnamefont {P.}~\bibnamefont {Bordia}},
  \bibinfo {author} {\bibfnamefont {H.~P.}\ \bibnamefont {L{\"u}schen}},
  \bibinfo {author} {\bibfnamefont {M.~H.}\ \bibnamefont {Fischer}}, \bibinfo
  {author} {\bibfnamefont {R.}~\bibnamefont {Vosk}}, \bibinfo {author}
  {\bibfnamefont {E.}~\bibnamefont {Altman}}, \bibinfo {author} {\bibfnamefont
  {U.}~\bibnamefont {Schneider}}, \ and\ \bibinfo {author} {\bibfnamefont
  {I.}~\bibnamefont {Bloch}},\ }\bibfield  {title} {\enquote {\bibinfo {title}
  {Observation of many-body localization of interacting fermions in a
  quasirandom optical lattice},}\ }\href {\doibase 10.1126/science.aaa7432}
  {\bibfield  {journal} {\bibinfo  {journal} {Science}\ }\textbf {\bibinfo
  {volume} {349}},\ \bibinfo {pages} {842} (\bibinfo {year}
  {2015})}\BibitemShut {NoStop}%
\bibitem [{\citenamefont {Bordia}\ \emph {et~al.}(2016)\citenamefont {Bordia},
  \citenamefont {L\"uschen}, \citenamefont {Hodgman}, \citenamefont
  {Schreiber}, \citenamefont {Bloch},\ and\ \citenamefont {Schneider}}]{bor16}%
  \BibitemOpen
  \bibfield  {author} {\bibinfo {author} {\bibfnamefont {P.}~\bibnamefont
  {Bordia}}, \bibinfo {author} {\bibfnamefont {H.~P.}\ \bibnamefont
  {L\"uschen}}, \bibinfo {author} {\bibfnamefont {S.~S.}\ \bibnamefont
  {Hodgman}}, \bibinfo {author} {\bibfnamefont {M.}~\bibnamefont {Schreiber}},
  \bibinfo {author} {\bibfnamefont {I.}~\bibnamefont {Bloch}}, \ and\ \bibinfo
  {author} {\bibfnamefont {U.}~\bibnamefont {Schneider}},\ }\bibfield  {title}
  {\enquote {\bibinfo {title} {Coupling identical one-dimensional many-body
  localized systems},}\ }\href {\doibase 10.1103/PhysRevLett.116.140401}
  {\bibfield  {journal} {\bibinfo  {journal} {Phys. Rev. Lett.}\ }\textbf
  {\bibinfo {volume} {116}},\ \bibinfo {pages} {140401} (\bibinfo {year}
  {2016})}\BibitemShut {NoStop}%
\bibitem [{\citenamefont {Choi}\ \emph {et~al.}(2016)\citenamefont {Choi},
  \citenamefont {Hild}, \citenamefont {Zeiher}, \citenamefont {Schau{\ss}},
  \citenamefont {Rubio-Abadal}, \citenamefont {Yefsah}, \citenamefont
  {Khemani}, \citenamefont {Huse}, \citenamefont {Bloch},\ and\ \citenamefont
  {Gross}}]{cho16}%
  \BibitemOpen
  \bibfield  {author} {\bibinfo {author} {\bibfnamefont {J.-Y.}\ \bibnamefont
  {Choi}}, \bibinfo {author} {\bibfnamefont {S.}~\bibnamefont {Hild}}, \bibinfo
  {author} {\bibfnamefont {J.}~\bibnamefont {Zeiher}}, \bibinfo {author}
  {\bibfnamefont {P.}~\bibnamefont {Schau{\ss}}}, \bibinfo {author}
  {\bibfnamefont {A.}~\bibnamefont {Rubio-Abadal}}, \bibinfo {author}
  {\bibfnamefont {T.}~\bibnamefont {Yefsah}}, \bibinfo {author} {\bibfnamefont
  {V.}~\bibnamefont {Khemani}}, \bibinfo {author} {\bibfnamefont {D.~A.}\
  \bibnamefont {Huse}}, \bibinfo {author} {\bibfnamefont {I.}~\bibnamefont
  {Bloch}}, \ and\ \bibinfo {author} {\bibfnamefont {C.}~\bibnamefont
  {Gross}},\ }\bibfield  {title} {\enquote {\bibinfo {title} {Exploring the
  many-body localization transition in two dimensions},}\ }\href {\doibase
  10.1126/science.aaf8834} {\bibfield  {journal} {\bibinfo  {journal}
  {Science}\ }\textbf {\bibinfo {volume} {352}},\ \bibinfo {pages} {1547}
  (\bibinfo {year} {2016})}\BibitemShut {NoStop}%
\bibitem [{\citenamefont {Huang}(2017)}]{hua17}%
  \BibitemOpen
  \bibfield  {author} {\bibinfo {author} {\bibfnamefont {Y.}~\bibnamefont
  {Huang}},\ }\bibfield  {title} {\enquote {\bibinfo {title} {Entanglement
  dynamics in critical random quantum {Ising} chain with perturbations},}\
  }\href {\doibase https://doi.org/10.1016/j.aop.2017.02.018} {\bibfield
  {journal} {\bibinfo  {journal} {Ann. Phys. (N. Y.)}\ }\textbf {\bibinfo
  {volume} {380}},\ \bibinfo {pages} {224} (\bibinfo {year}
  {2017})}\BibitemShut {NoStop}%
\bibitem [{\citenamefont {De~Roeck}\ and\ \citenamefont
  {Huveneers}(2017)}]{der17}%
  \BibitemOpen
  \bibfield  {author} {\bibinfo {author} {\bibfnamefont {W.}~\bibnamefont
  {De~Roeck}}\ and\ \bibinfo {author} {\bibfnamefont {F.}~\bibnamefont
  {Huveneers}},\ }\bibfield  {title} {\enquote {\bibinfo {title} {Stability and
  instability towards delocalization in many-body localization systems},}\
  }\href {\doibase 10.1103/PhysRevB.95.155129} {\bibfield  {journal} {\bibinfo
  {journal} {Phys. Rev. B}\ }\textbf {\bibinfo {volume} {95}},\ \bibinfo
  {pages} {155129} (\bibinfo {year} {2017})}\BibitemShut {NoStop}%
\bibitem [{\citenamefont {Papić}\ \emph {et~al.}(2015)\citenamefont {Papić},
  \citenamefont {Stoudenmire},\ and\ \citenamefont {Abanin}}]{pap15}%
  \BibitemOpen
  \bibfield  {author} {\bibinfo {author} {\bibfnamefont {Z.}~\bibnamefont
  {Papić}}, \bibinfo {author} {\bibfnamefont {E.~M.}\ \bibnamefont
  {Stoudenmire}}, \ and\ \bibinfo {author} {\bibfnamefont {D.~A.}\ \bibnamefont
  {Abanin}},\ }\bibfield  {title} {\enquote {\bibinfo {title} {Many-body
  localization in disorder-free systems: The importance of finite-size
  constraints},}\ }\href {\doibase https://doi.org/10.1016/j.aop.2015.08.024}
  {\bibfield  {journal} {\bibinfo  {journal} {Ann. Phys. (N. Y.)}\ }\textbf
  {\bibinfo {volume} {362}},\ \bibinfo {pages} {714 -- 725} (\bibinfo {year}
  {2015})}\BibitemShut {NoStop}%
\bibitem [{\citenamefont {Yao}\ \emph {et~al.}(2016)\citenamefont {Yao},
  \citenamefont {Laumann}, \citenamefont {Cirac}, \citenamefont {Lukin},\ and\
  \citenamefont {Moore}}]{yao16}%
  \BibitemOpen
  \bibfield  {author} {\bibinfo {author} {\bibfnamefont {N.~Y.}\ \bibnamefont
  {Yao}}, \bibinfo {author} {\bibfnamefont {C.~R.}\ \bibnamefont {Laumann}},
  \bibinfo {author} {\bibfnamefont {J.~I.}\ \bibnamefont {Cirac}}, \bibinfo
  {author} {\bibfnamefont {M.~D.}\ \bibnamefont {Lukin}}, \ and\ \bibinfo
  {author} {\bibfnamefont {J.~E.}\ \bibnamefont {Moore}},\ }\bibfield  {title}
  {\enquote {\bibinfo {title} {Quasi-many-body localization in
  translation-invariant systems},}\ }\href {\doibase
  10.1103/PhysRevLett.117.240601} {\bibfield  {journal} {\bibinfo  {journal}
  {Phys. Rev. Lett.}\ }\textbf {\bibinfo {volume} {117}},\ \bibinfo {pages}
  {240601} (\bibinfo {year} {2016})}\BibitemShut {NoStop}%
\bibitem [{\citenamefont {Bera}\ \emph {et~al.}(2017)\citenamefont {Bera},
  \citenamefont {De~Tomasi}, \citenamefont {Weiner},\ and\ \citenamefont
  {Evers}}]{sou17}%
  \BibitemOpen
  \bibfield  {author} {\bibinfo {author} {\bibfnamefont {Soumya}\ \bibnamefont
  {Bera}}, \bibinfo {author} {\bibfnamefont {Giuseppe}\ \bibnamefont
  {De~Tomasi}}, \bibinfo {author} {\bibfnamefont {Felix}\ \bibnamefont
  {Weiner}}, \ and\ \bibinfo {author} {\bibfnamefont {Ferdinand}\ \bibnamefont
  {Evers}},\ }\bibfield  {title} {\enquote {\bibinfo {title} {Density
  propagator for many-body localization: Finite-size effects, transient
  subdiffusion, and exponential decay},}\ }\href {\doibase
  10.1103/PhysRevLett.118.196801} {\bibfield  {journal} {\bibinfo  {journal}
  {Phys. Rev. Lett.}\ }\textbf {\bibinfo {volume} {118}},\ \bibinfo {pages}
  {196801} (\bibinfo {year} {2017})}\BibitemShut {NoStop}%
\bibitem [{\citenamefont {Smith}\ \emph
  {et~al.}(2017{\natexlab{a}})\citenamefont {Smith}, \citenamefont {Knolle},
  \citenamefont {Kovrizhin},\ and\ \citenamefont {Moessner}}]{smi17a}%
  \BibitemOpen
  \bibfield  {author} {\bibinfo {author} {\bibfnamefont {A.}~\bibnamefont
  {Smith}}, \bibinfo {author} {\bibfnamefont {J.}~\bibnamefont {Knolle}},
  \bibinfo {author} {\bibfnamefont {D.~L.}\ \bibnamefont {Kovrizhin}}, \ and\
  \bibinfo {author} {\bibfnamefont {R.}~\bibnamefont {Moessner}},\ }\bibfield
  {title} {\enquote {\bibinfo {title} {Disorder-free localization},}\ }\href
  {\doibase 10.1103/PhysRevLett.118.266601} {\bibfield  {journal} {\bibinfo
  {journal} {Phys. Rev. Lett.}\ }\textbf {\bibinfo {volume} {118}},\ \bibinfo
  {pages} {266601} (\bibinfo {year} {2017}{\natexlab{a}})}\BibitemShut
  {NoStop}%
\bibitem [{\citenamefont {Smith}\ \emph
  {et~al.}(2017{\natexlab{b}})\citenamefont {Smith}, \citenamefont {Knolle},
  \citenamefont {Moessner},\ and\ \citenamefont {Kovrizhin}}]{smi17b}%
  \BibitemOpen
  \bibfield  {author} {\bibinfo {author} {\bibfnamefont {A.}~\bibnamefont
  {Smith}}, \bibinfo {author} {\bibfnamefont {J.}~\bibnamefont {Knolle}},
  \bibinfo {author} {\bibfnamefont {R.}~\bibnamefont {Moessner}}, \ and\
  \bibinfo {author} {\bibfnamefont {D.~L.}\ \bibnamefont {Kovrizhin}},\
  }\bibfield  {title} {\enquote {\bibinfo {title} {Absence of ergodicity
  without quenched disorder: From quantum disentangled liquids to many-body
  localization},}\ }\href {\doibase 10.1103/PhysRevLett.119.176601} {\bibfield
  {journal} {\bibinfo  {journal} {Phys. Rev. Lett.}\ }\textbf {\bibinfo
  {volume} {119}},\ \bibinfo {pages} {176601} (\bibinfo {year}
  {2017}{\natexlab{b}})}\BibitemShut {NoStop}%
\bibitem [{\citenamefont {Marcuzzi}\ \emph {et~al.}(2013)\citenamefont
  {Marcuzzi}, \citenamefont {Marino}, \citenamefont {Gambassi},\ and\
  \citenamefont {Silva}}]{mar13}%
  \BibitemOpen
  \bibfield  {author} {\bibinfo {author} {\bibfnamefont {M.}~\bibnamefont
  {Marcuzzi}}, \bibinfo {author} {\bibfnamefont {J.}~\bibnamefont {Marino}},
  \bibinfo {author} {\bibfnamefont {A.}~\bibnamefont {Gambassi}}, \ and\
  \bibinfo {author} {\bibfnamefont {A.}~\bibnamefont {Silva}},\ }\bibfield
  {title} {\enquote {\bibinfo {title} {Prethermalization in a nonintegrable
  quantum spin chain after a quench},}\ }\href {\doibase
  10.1103/PhysRevLett.111.197203} {\bibfield  {journal} {\bibinfo  {journal}
  {Phys. Rev. Lett.}\ }\textbf {\bibinfo {volume} {111}},\ \bibinfo {pages}
  {197203} (\bibinfo {year} {2013})}\BibitemShut {NoStop}%
\bibitem [{\citenamefont {Essler}\ \emph {et~al.}(2014)\citenamefont {Essler},
  \citenamefont {Kehrein}, \citenamefont {Manmana},\ and\ \citenamefont
  {Robinson}}]{ess14}%
  \BibitemOpen
  \bibfield  {author} {\bibinfo {author} {\bibfnamefont {F.~H.~L.}\
  \bibnamefont {Essler}}, \bibinfo {author} {\bibfnamefont {S.}~\bibnamefont
  {Kehrein}}, \bibinfo {author} {\bibfnamefont {S.~R.}\ \bibnamefont
  {Manmana}}, \ and\ \bibinfo {author} {\bibfnamefont {N.~J.}\ \bibnamefont
  {Robinson}},\ }\bibfield  {title} {\enquote {\bibinfo {title} {Quench
  dynamics in a model with tuneable integrability breaking},}\ }\href {\doibase
  10.1103/PhysRevB.89.165104} {\bibfield  {journal} {\bibinfo  {journal} {Phys.
  Rev. B}\ }\textbf {\bibinfo {volume} {89}},\ \bibinfo {pages} {165104}
  (\bibinfo {year} {2014})}\BibitemShut {NoStop}%
\bibitem [{\citenamefont {Bertini}\ \emph {et~al.}(2015)\citenamefont
  {Bertini}, \citenamefont {Essler}, \citenamefont {Groha},\ and\ \citenamefont
  {Robinson}}]{ber15}%
  \BibitemOpen
  \bibfield  {author} {\bibinfo {author} {\bibfnamefont {B.}~\bibnamefont
  {Bertini}}, \bibinfo {author} {\bibfnamefont {F.~H.~L.}\ \bibnamefont
  {Essler}}, \bibinfo {author} {\bibfnamefont {S.}~\bibnamefont {Groha}}, \
  and\ \bibinfo {author} {\bibfnamefont {N.~J.}\ \bibnamefont {Robinson}},\
  }\bibfield  {title} {\enquote {\bibinfo {title} {Prethermalization and
  thermalization in models with weak integrability breaking},}\ }\href
  {\doibase 10.1103/PhysRevLett.115.180601} {\bibfield  {journal} {\bibinfo
  {journal} {Phys. Rev. Lett.}\ }\textbf {\bibinfo {volume} {115}},\ \bibinfo
  {pages} {180601} (\bibinfo {year} {2015})}\BibitemShut {NoStop}%
\bibitem [{\citenamefont {Lan}\ \emph {et~al.}(2018)\citenamefont {Lan},
  \citenamefont {van Horssen}, \citenamefont {Powell},\ and\ \citenamefont
  {Garrahan}}]{lan18}%
  \BibitemOpen
  \bibfield  {author} {\bibinfo {author} {\bibfnamefont {Z.}~\bibnamefont
  {Lan}}, \bibinfo {author} {\bibfnamefont {M.}~\bibnamefont {van Horssen}},
  \bibinfo {author} {\bibfnamefont {S.}~\bibnamefont {Powell}}, \ and\ \bibinfo
  {author} {\bibfnamefont {J.~P.}\ \bibnamefont {Garrahan}},\ }\bibfield
  {title} {\enquote {\bibinfo {title} {Quantum slow relaxation and
  metastability due to dynamical constraints},}\ }\href {\doibase
  10.1103/PhysRevLett.121.040603} {\bibfield  {journal} {\bibinfo  {journal}
  {Phys. Rev. Lett.}\ }\textbf {\bibinfo {volume} {121}},\ \bibinfo {pages}
  {040603} (\bibinfo {year} {2018})}\BibitemShut {NoStop}%
\bibitem [{\citenamefont {Kohmoto}\ \emph {et~al.}(1981)\citenamefont
  {Kohmoto}, \citenamefont {den Nijs},\ and\ \citenamefont {Kadanoff}}]{koh81}%
  \BibitemOpen
  \bibfield  {author} {\bibinfo {author} {\bibfnamefont {M.}~\bibnamefont
  {Kohmoto}}, \bibinfo {author} {\bibfnamefont {M.}~\bibnamefont {den Nijs}}, \
  and\ \bibinfo {author} {\bibfnamefont {L.~P.}\ \bibnamefont {Kadanoff}},\
  }\bibfield  {title} {\enquote {\bibinfo {title} {Hamiltonian studies of the
  $d=2$ ashkin-teller model},}\ }\href {\doibase 10.1103/PhysRevB.24.5229}
  {\bibfield  {journal} {\bibinfo  {journal} {Phys. Rev. B}\ }\textbf {\bibinfo
  {volume} {24}},\ \bibinfo {pages} {5229} (\bibinfo {year}
  {1981})}\BibitemShut {NoStop}%
\bibitem [{\citenamefont {Atas}\ \emph {et~al.}(2013)\citenamefont {Atas},
  \citenamefont {Bogomolny}, \citenamefont {Giraud},\ and\ \citenamefont
  {Roux}}]{ata13}%
  \BibitemOpen
  \bibfield  {author} {\bibinfo {author} {\bibfnamefont {Y.~Y.}\ \bibnamefont
  {Atas}}, \bibinfo {author} {\bibfnamefont {E.}~\bibnamefont {Bogomolny}},
  \bibinfo {author} {\bibfnamefont {O.}~\bibnamefont {Giraud}}, \ and\ \bibinfo
  {author} {\bibfnamefont {G.}~\bibnamefont {Roux}},\ }\bibfield  {title}
  {\enquote {\bibinfo {title} {Distribution of the ratio of consecutive level
  spacings in random matrix ensembles},}\ }\href {\doibase
  10.1103/PhysRevLett.110.084101} {\bibfield  {journal} {\bibinfo  {journal}
  {Phys. Rev. Lett.}\ }\textbf {\bibinfo {volume} {110}},\ \bibinfo {pages}
  {084101} (\bibinfo {year} {2013})}\BibitemShut {NoStop}%
\bibitem [{\citenamefont {Madhukar}\ and\ \citenamefont {Post}(1977)}]{mad77}%
  \BibitemOpen
  \bibfield  {author} {\bibinfo {author} {\bibfnamefont {A.}~\bibnamefont
  {Madhukar}}\ and\ \bibinfo {author} {\bibfnamefont {W.}~\bibnamefont
  {Post}},\ }\bibfield  {title} {\enquote {\bibinfo {title} {Exact solution for
  the diffusion of a particle in a medium with site diagonal and off-diagonal
  dynamic disorder},}\ }\href {\doibase 10.1103/PhysRevLett.39.1424} {\bibfield
   {journal} {\bibinfo  {journal} {Phys. Rev. Lett.}\ }\textbf {\bibinfo
  {volume} {39}},\ \bibinfo {pages} {1424} (\bibinfo {year}
  {1977})}\BibitemShut {NoStop}%
\bibitem [{\citenamefont {Shiraishi}\ and\ \citenamefont {Mori}(2017)}]{shi17}%
  \BibitemOpen
  \bibfield  {author} {\bibinfo {author} {\bibfnamefont {N.}~\bibnamefont
  {Shiraishi}}\ and\ \bibinfo {author} {\bibfnamefont {T.}~\bibnamefont
  {Mori}},\ }\bibfield  {title} {\enquote {\bibinfo {title} {Systematic
  construction of counterexamples to the eigenstate thermalization
  hypothesis},}\ }\href {\doibase 10.1103/PhysRevLett.119.030601} {\bibfield
  {journal} {\bibinfo  {journal} {Phys. Rev. Lett.}\ }\textbf {\bibinfo
  {volume} {119}},\ \bibinfo {pages} {030601} (\bibinfo {year}
  {2017})}\BibitemShut {NoStop}%
\bibitem [{\citenamefont {Turner}\ \emph {et~al.}(2018)\citenamefont {Turner},
  \citenamefont {Michailidis}, \citenamefont {Abanin}, \citenamefont {Serbyn},\
  and\ \citenamefont {Papić}}]{tur18}%
  \BibitemOpen
  \bibfield  {author} {\bibinfo {author} {\bibfnamefont {C.~J.}\ \bibnamefont
  {Turner}}, \bibinfo {author} {\bibfnamefont {A.~A.}\ \bibnamefont
  {Michailidis}}, \bibinfo {author} {\bibfnamefont {D.~A.}\ \bibnamefont
  {Abanin}}, \bibinfo {author} {\bibfnamefont {M.}~\bibnamefont {Serbyn}}, \
  and\ \bibinfo {author} {\bibfnamefont {Z.}~\bibnamefont {Papić}},\
  }\bibfield  {title} {\enquote {\bibinfo {title} {Weak ergodicity breaking
  from quantum many-body scars},}\ }\href@noop {} {\bibfield  {journal}
  {\bibinfo  {journal} {Nature Physics}\ }\textbf {\bibinfo {volume} {14}},\
  \bibinfo {pages} {745} (\bibinfo {year} {2018})}\BibitemShut {NoStop}%
\bibitem [{\citenamefont {Pai}\ \emph {et~al.}(2019)\citenamefont {Pai},
  \citenamefont {Pretko},\ and\ \citenamefont {Nandkishore}}]{pai19}%
  \BibitemOpen
  \bibfield  {author} {\bibinfo {author} {\bibfnamefont {S.}~\bibnamefont
  {Pai}}, \bibinfo {author} {\bibfnamefont {M.}~\bibnamefont {Pretko}}, \ and\
  \bibinfo {author} {\bibfnamefont {R.~M.}\ \bibnamefont {Nandkishore}},\
  }\bibfield  {title} {\enquote {\bibinfo {title} {Localization in fractonic
  random circuits},}\ }\href {\doibase 10.1103/PhysRevX.9.021003} {\bibfield
  {journal} {\bibinfo  {journal} {Phys. Rev. X}\ }\textbf {\bibinfo {volume}
  {9}},\ \bibinfo {pages} {021003} (\bibinfo {year} {2019})}\BibitemShut
  {NoStop}%
\bibitem [{\citenamefont {Sala}\ \emph {et~al.}()\citenamefont {Sala},
  \citenamefont {Rakovszky}, \citenamefont {Verresen}, \citenamefont {Knap},\
  and\ \citenamefont {Pollmann}}]{sal19}%
  \BibitemOpen
  \bibfield  {author} {\bibinfo {author} {\bibfnamefont {P.}~\bibnamefont
  {Sala}}, \bibinfo {author} {\bibfnamefont {T.}~\bibnamefont {Rakovszky}},
  \bibinfo {author} {\bibfnamefont {R.}~\bibnamefont {Verresen}}, \bibinfo
  {author} {\bibfnamefont {M.}~\bibnamefont {Knap}}, \ and\ \bibinfo {author}
  {\bibfnamefont {F.}~\bibnamefont {Pollmann}},\ }\href@noop {} {\enquote
  {\bibinfo {title} {Ergodicity-breaking arising from {Hilbert} space
  fragmentation in dipole-conserving {Hamiltonians}},}\ }\Eprint
  {http://arxiv.org/abs/arXiv:1904.04266} {arXiv:1904.04266} \BibitemShut
  {NoStop}%
\bibitem [{\citenamefont {Khemani}\ and\ \citenamefont
  {Nandkishore}()}]{khe19}%
  \BibitemOpen
  \bibfield  {author} {\bibinfo {author} {\bibfnamefont {V.}~\bibnamefont
  {Khemani}}\ and\ \bibinfo {author} {\bibfnamefont {R.}~\bibnamefont
  {Nandkishore}},\ }\href@noop {} {\enquote {\bibinfo {title} {Local
  constraints can globally shatter hilbert space: a new route to quantum
  information protection},}\ }\Eprint {http://arxiv.org/abs/arXiv:1904.04815}
  {arXiv:1904.04815} \BibitemShut {NoStop}%
\end{thebibliography}%

\end{document}